\documentclass[secnumarabic, graphics,floatfix, nofootinbib,tightenlines,nobibnotes, aps, prl, 12pt]{revtex4-1}
\usepackage{graphicx}
\usepackage[english]{babel}
\usepackage[utf8]{inputenc}
\usepackage{amsmath,amssymb}
\usepackage{mathrsfs}
\usepackage{amsfonts}
\usepackage{multirow}
\usepackage{pstricks}
\usepackage{float}
\usepackage{subfigure}
\usepackage{color}
\usepackage{epsfig}
\usepackage{pst-tools}

\begin{document}
 \newcommand{\bq}{\begin{equation}}
 \newcommand{\eq}{\end{equation}}
 \newcommand{\bqn}{\begin{eqnarray}}
 \newcommand{\eqn}{\end{eqnarray}}
 \newcommand{\nb}{\nonumber}
 \newcommand{\lb}{\label}
 
\title{On quasinormal frequencies of black hole perturbations with an external source}

\author{Wei-Liang Qian$^{1,2,3}$}
\author{Kai Lin $^{4, 2}$}
\author{Jian-Pin Wu$^{1}$}
\author{Bin Wang$^{1}$}
\author{Rui-Hong Yue$^{1}$}

\affiliation{$^{1}$ Center for Gravitation and Cosmology, College of Physical Science and Technology, Yangzhou University, 225009, Yangzhou, China}
\affiliation{$^{2}$ Institute for theoretical physics and cosmology, Zhejiang University of Technology, 310032, Hangzhou, China}
\affiliation{$^{3}$ Escola de Engenharia de Lorena, Universidade de S\~ao Paulo, 12602-810, Lorena, SP, Brazil}
\affiliation{$^{4}$ Hubei Subsurface Multi-scale Imaging Key Laboratory, Institute of Geophysics and Geomatics, China University of Geosciences, 430074, Wuhan, Hubei, China}

\date{Sept. 25th, 2020}

\begin{abstract}

In the study of perturbations around black hole configurations, whether an external source can influence the perturbation behavior is an interesting topic to investigate. 
When the source acts as an initial pulse, it is intuitively acceptable that the existing quasinormal frequencies will remain unchanged. 
However, the confirmation of such an intuition is not trivial for the rotating black hole, since the eigenvalues in the radial and angular parts of the master equations are coupled. 
We show that for the rotating black holes, a moderate source term in the master equation in the Laplace s-domain does not modify the quasinormal modes.
Furthermore, we generalize our discussions to the case where the external source serves as a driving force. 
Different from an initial pulse, an external source may further drive the system to experience new perturbation modes. 
To be specific, novel dissipative singularities might be brought into existence and enrich the pole structure. 
This is a physically relevant scenario, due to its possible implication in modified gravity.
Our arguments are based on exploring the pole structure of the solution in the Laplace s-domain with the presence of the external source.
The analytical analyses are verified numerically by solving the inhomogeneous differential equation and extracting the dominant complex frequencies by employing the Prony method.

\end{abstract}

\maketitle

\section{I. Introduction}

The physical content regarding perturbations in a black hole spacetime can be viewed as reminiscent of a damped harmonic oscillator.
Due to the dissipative nature of the system, the frequencies of the oscillation are usually complex.
It is well-known that for a harmonic oscillator, its natural frequencies are independent of the specific initial pulse.
However, when a sinusoidal driving force is applied, quite the contrary, the frequency of the steady-state solution is governed by that of the external force.
This indicates the distinct characteristics between the initial pulse and the external driving force.
Not surprisingly, these concepts can be explored analogously in the context of black hole perturbations. 
In fact, the problem of black hole quasinormal mode~\cite{agr-qnm-review-01,agr-qnm-review-02,agr-qnm-review-06,agr-qnm-review-03,agr-qnm-review-04} is more sophisticated.
As an open system, the dissipation demonstrates itself by ingoing waves at the horizon or the outgoing waves at infinity in asymptotically flat spacetimes, which subsequently leads to energy loss.
Subsequently, for a non-Hermitian system, the relevant excited states are those of quasinormal modes with complex frequencies.
Besides, the boundary condition demands more strenuous efforts, as the solution diverges at both spatial boundaries.

For black hole quasinormal modes, most studies concern the master equation without any explicit external source in the time-domain.
In other words, the master equation in the time-domain is a homogeneous equation, where the initial perturbation pulse furnishes the system with an initial condition.
In the Laplace s-domain, however, the initial condition is transformed to the r.h.s. of the equation, so that the resultant ordinary second-order differential equation becomes inhomogeneous.
Nonetheless, one intuitively argues that the resultant source term in the s-domain is of little physical relevance, as it does not affect the existing quasinormal modes.
Also, we note that the above scenario is largely related to the matter being minimally coupled to the curvature in Einstein's general relativity in spherical symmetry. 
In the Kerr black hole background, the radial part of the master equation is not a single second-order differential equation. 
Its eigenvalue is coupled to that of the angular part of the master equation, and therefore, it is not obvious why the initial pulse will not influence the quasinormal frequencies of the Kerr black holes. 
Further investigation is called for. 

Moreover, motivated mainly by the observed accelerated cosmic expansion, theories of modified gravity have become a topic of increasing interest in the last decades. 
Among many promising possibilities, the latter includes scalar-tensor~\cite{agr-modified-gravity-Horndeski-01,agr-modified-gravity-dhost-01}, vector-tensor~\cite{agr-modified-gravity-vector-01, agr-modified-gravity-vector-02}, and scalar-vector-tensor~\cite{agr-modified-Moffat-01} theories.
There, the matter field can be non-minimally coupled to the curvature sector, and therefore, one might expect the resultant master equations in the time-domain to become inhomogeneous.
As an example, the degenerate higher-order scalar-tensor (DHOST) theory~\cite{agr-modified-gravity-dhost-01, agr-modified-gravity-dhost-02, agr-modified-gravity-dhost-03} may admit ``stealth" solutions~\cite{agr-modified-gravity-dhost-06, agr-modified-gravity-dhost-10, agr-modified-gravity-dhost-11, agr-modified-gravity-dhost-12}.
The latter does not influence the background metric due to its vanishing energy-momentum tensor~\cite{agr-qnm-17}.
The resultant metric differs from the Kerr one by dressing up with a linearly-time dependent field.
Indeed, as it has been demonstrated~\cite{agr-bh-nohair-03,agr-bh-nohair-04}, under moderate hypotheses, the only non-trivial modification that can be obtained is at the perturbation level. 
In this regard, the metric perturbations in the DHOST theories have been investigated ~\cite{agr-modified-gravity-dhost-07, agr-modified-gravity-dhost-08,  agr-modified-gravity-dhost-09} recently.
It was shown that the equation of motion for the tensorial perturbations are characterized by some intriguing features.
To be specific, the scalar perturbation is shown to be decoupled from those of the Einstein tensor.
This result leads to immediate simplifications, namely, the time-domain master equations for the tensor perturbations possess the form of linearized Einstein equations supplemented with a source term.
The latter, in turn, is governed by the scalar perturbation.
Therefore, it is natural to expect that the study of the related quasinormal modes may provide essential information on the stealth scalar hair, as well as the properties of the spacetime of the gravity theory in question.
Furthermore, as an external driving force affects the harmonic oscillator, the external source is expected to trigger novelty.
One physically relevant example related to the external field source is a {\it quench}, introduced to act as a driving force to influence the system~\cite{condensed-quench-02, condensed-quench-review-01}.
For instance, a holographic analysis of quench is carried out in Ref.~\cite{adscft-condensed-quench-06}, where a zero mode, reminiscent of the Kibble-Zurek scaling in the dual system, was disclosed.
In particular, the specific mode does not belong to the original metric neither the gauge field, it is obtained via a time-dependent source introduced onto the boundary.

The present work is motivated by the above intriguing scenarios and aims to study the properties of quasinormal modes with external sources in the time-domain.
To be specific, we will investigate the case when the master equation in the time-domain is inhomogeneous.
In order to show the influences in perturbation behaviors caused by different characteristics of the source, we first concentrate our attention on the case where the source takes on the role of an initial pulse.  
Intuitively the quasinormal frequencies should not be affected by such an initial pulse source.
We will confirm that such intuition holds not only for static spherical black hole backgrounds but also for rotating configurations. 
It is especially not straightforward to confirm such intuition for rotating case, since the eigenvalues of the radial and angular parts of the master equation are coupled, which makes the original arguments based on contour integral invalid. 
Moreover, in comparison to the case of a driven harmonic oscillator, it is meaningful to examine further the non-trivial influences on the perturbation caused by external source terms. 
We will show that different from the initial pulse source, the external field introduced to the system may bring additional modes to the perturbed system. 
This is because the external source term can introduce dissipative singularities in the complex plane, which results in novel modes in the system.

The organization of the paper is as follows.
In the following section, we first discuss the solution of a specific driven harmonic oscillator. 
Then in section III, we generalize to a rigorous discussion on black hole quasinormal models in terms of the analysis of the pole structure of the associated Green function in the Laplace s-domain. 
We concentrate on the effects of the source on the perturbated system as the initial pulse.  
In section III.A, we confirm the physical intuition on the influence of the initial pulse in the perturbation around static spherical black hole backgrounds. 
In section III.B, we present a proof to support the intuition on the initial pulse source effect on perturbation system in rotating configurations.
In section IV, we extend our discussions to the external source influence on the perturbation system generated by the external field. 
We show that the external source term may induce additional quasinormal frequencies due to its modifications to the pole structures of the solution. 
We present numerical confirmation to support our analytic arguments in section V.
The last section is devoted to further discussions and conclusions.

\section{II. The quasinormal frequencies of a vibrating string with dissipation}

As characterized by complex natural frequencies, a damped oscillator is usually employed to illustrate the physical content for the quasinormal oscillations in a dissipation system.
Moreover, regarding a dissipative wave equation, a vibrating string subjected to a driven force is an appropriate analogy.
To illustrate the main idea, the following derivations concerns a toy model investigated recently by the authors of Ref.~\cite{agr-modified-gravity-dhost-08}.
In this model, the wave propagating along the string is governed by the following dissipative wave equation with a source term, namely,
\begin{eqnarray}
\frac{\partial^2 \Psi}{\partial x^2}-\frac{\partial^2\Psi}{\partial t^2}-\frac{2}{\tau}\frac{\partial\Psi}{\partial t}=S(t, x) ,
\label{toy_eq}
\end{eqnarray}
where the the string is held fixed at both ends, and therefore the wave function $\Psi(t, x)$, as well as the source $S(t, x)$, satisfies the boundary conditions 
\begin{eqnarray}
\Psi(t, 0)=\Psi(t, L)=0 \nonumber\\
S(t, 0)=S(t, L)=0,
\label{toy_boundary_conditions}
\end{eqnarray}
respectively, where $L$ denotes the length of the string.
Here, the relaxation time $\tau = \mathrm{const.}$ carries out the role of a simplified dissipation mechanism.

If the source term vanishes, $\Psi$ is governed by the superposition of the quasinormal oscillations
\begin{eqnarray}
\Psi(t, x) = \sum_n A_n \sin(n\pi x/L)e^{-i\omega_n t} ,
\label{toy_sol_sourceless}
\end{eqnarray}
with complex frequencies $\omega_n$ given by
\begin{eqnarray}
\omega_n^\pm \tau \equiv -i\pm \sqrt{n^2\frac{\pi^2\tau^2}{L^2}-1} .
\label{toy_qnm}
\end{eqnarray}

With the presence of the source term, it is straightforward to verify that the formal solution of the wave equation Eq.~\eqref{toy_eq} reads
\begin{eqnarray}
\Psi(t, x)=-\sum_n\int_{-\infty}^\infty d\omega\frac{\mathscr{S}(\omega, n)\sin(n\pi x/L)e^{-i\omega t}}{n^2\pi^2/L^2-\omega^2-2i\omega/\tau} ,
\label{toy_sol_source_formal}
\end{eqnarray}
where we have conveniently expanded the source term $S(t, x)$ in the form
\begin{eqnarray}
S(t, x)=\sum_n\int_{-\infty}^\infty d\omega \sin(n\pi x/L) e^{-i\omega t}\mathscr{S}(\omega, n).
\label{toy_source}
\end{eqnarray}

We note that the integral range for the variable $\omega$ is from $-\infty$ to $\infty$, as it originates from a Fourier transform.
Subsequently, for $t>0$, according to Jordan's lemma, one chooses the contour to close the integral around the lower half-plane.
As a result, the integral evaluates to the summation of the residue of the integrant, namely,
\begin{eqnarray}
\Psi(t, x)&=&2\pi i\sum_n \mathrm{Res} \left(\frac{\mathscr{S}(\omega, n)\sin(n\pi x/L)e^{-i\omega t}}{n^2\pi^2/L^2-\omega^2-2i\omega/\tau}\right)\nonumber\\
&=& 2\pi i\sum_n  \frac{\sin(n\pi x/L)}{\omega_n^+-\omega_n^-}\left[\mathscr{S}(\omega_n^+, n)e^{-i\omega_n^+ t}-\mathscr{S}(\omega_n^-, n)e^{-i\omega_n^- t}\right].
\label{toy_sol_source_res}
\end{eqnarray}
Here we have assumed that the source term is a moderately benign function in the sense that $\mathscr{S}(\omega, n)$ does not contain any singularity.
The quasinormal frequencies can be read off by examing the temporal dependence of the above results, namely, the $e^{i\omega_n^\pm t}$ factors.
They are, therefore, determined by the residues of the integrant.
The latter is related to the zeros of the denominator on the first line of Eq.~\eqref{toy_sol_source_res}, which are precisely the frequencies given in Eq.~\eqref{toy_qnm}.

It is observed that both quasinormal frequencies given by Eq.~\eqref{toy_qnm} are below the real axis, and therefore will be taken into account by the residue theorem.
If $n^2\pi^2\tau^2/L^2 > 1$, the two frequencies lie on a horizontal line, symmetric about the imaginary axis. 
If, on the other hand,  $n^2\pi^2\tau^2/L^2 < 1$, both frequencies lie on the imaginary axis below the origin.
It is rather interesting to point out that the above simple model also furnishes an elementary example of the so-called {\it gapped momentum states}~\cite{GMS-01, GMS-02}, where a gap is present in its dispersion relation but on the momentum axis.

Under moderate assumption for the source term, we have arrived at a conclusion which seemingly contradicts the example of driven harmonic oscillator given initially.
In the following section, we first extend the results to the context of black hole quasinormal modes.
Then, section IV, we resolve the above apparent contradiction by further exploring the different characteristic between the initial pulse and external driving force.

\section{III. Moderate external source acting as the initial pulse}

At this point, one might argue that the analogy given in the last section can only be viewed as a toy model when compared with the problem of black hole quasinormal modes.
First, the boundary condition of the problem is different: the solution is divergent at both spatial boundaries.
Besides, the system is dissipative not due to localized friction but owing to the energy loss from its boundary, namely, the ingoing waves at the horizon and/or the outgoing waves at infinity.
As a result, the oscillation frequencies are complex, which can be further traced to the fact that the system is non-Hermitian.
In terms of the wave equation, the term concerning the relaxation time $\tau$ is replaced by an effective potential.
Nonetheless, in the present section, we show that a similar conclusion can also be reached for black hole quasinormal modes.
We first discuss static black hole metric, then extend the results to the case of perturbations in rotating black holes.
The last subsection is devoted to the scenario where the external source itself introduces additional quasinormal frequencies.

\subsection{A. Schwarzschild black hole metric}

For a static black hole metric, the perturbation equation of various types of perturbations can be simplified by using the method of separation of variables $\chi=\Psi(t,r)S(\theta)e^{im\varphi}$.
The radial part of the master equation is a second order differential equation~\cite{agr-qnm-review-03,agr-qnm-review-06}.
\begin{eqnarray}
\frac{\partial^2}{\partial t^2}\Psi(t, x)+\left(-\frac{\partial^2}{\partial x^2}+V\right)\Psi(t, x)=0 ,
\label{master_eq_ns}
\end{eqnarray}
where the effective potential $V$ is determined by the given spacetime metric, spin ${\bar{s}}$ and angular momentum $\ell$ of the perturbation.
For instance, in four-dimensional Schwarzschild or SAdS metric, for massless scalar, electromagnetic and vector gravitational perturbations, it reads
\bqn
V=f\left[\frac{\ell(\ell+1)}{r^2}+(1-{\bar{s}}^2)\left(\frac{2M}{r^3}+\frac{4-{\bar{s}}^2}{2L^2}\right)\right] ,
\lb{V_master}
\eqn
where 
\bqn
f=1+r^2/L^2-2M/r ,
\lb{f_master}
\eqn
$M$ is the mass of the black hole, and $L$ represents the curvature radius of AdS spacetime so that the Schwarzschild geometry corresponds to $L\to\infty$.
The master equation is often conveniently expressed in terms of the tortoise coordinate, $x\equiv r_*(r)=\int dr/f$.
By expanding the external source in terms of spherical harmonics at a given radius, $r$ or $x$, the resultant radial equation is given by
\begin{eqnarray}
\frac{\partial^2}{\partial t^2}\Psi(t, x)+\left(-\frac{\partial^2}{\partial x^2}+V(x)\right)\Psi(t, x)=S(t, x) ,
\label{master_eq}
\end{eqnarray}
where $S(t, x)$ corresponds to the expansion coefficient for a given harmonics $(\ell,m)$.

In what follows, one employs the procedure by carrying out the Laplace transform in the time domain~\cite{agr-qnm-12,agr-qnm-review-01,agr-qnm-review-02}, 
\begin{eqnarray}
\hat{f}(s, x)=\int_0^\infty e^{-st} \Psi(t, x) dt \nonumber ,\\
\mathscr{S}(s, x)=\int_0^\infty e^{-st} S(t, x) dt ,
\label{master_Laplace}
\end{eqnarray}
Subsequently, the resultant radial equation in s-domain reads
\begin{eqnarray}
\hat{f}''(s, x)+\left(-s^2-V(x)\right)\hat{f}(s, x)=\mathcal{I}(s, x) - \mathscr{S}(s, x) .
\label{master_eq_s}
\end{eqnarray}
where a prime $'$ indicates the derivative regarding $x$, and the source terms on the r.h.s. of the equation consist of $\mathscr{S}(s, x)$ and $\mathcal{I}(s, x)$.
The latter is governed by the initial condition
\begin{eqnarray}
\mathcal{I}(s, x) = -s\left.\Psi\right|_{t=0} - \left.\frac{\partial \Psi}{\partial t}\right|_{t=0}.
\label{master_eq_sic}
\end{eqnarray}
We note that the lower limit of the integrations in Eqs.~\eqref{master_Laplace} is ``0".
Subsequently, $\hat{f}(s, x)$ and $\mathscr{S}(s, x)$ are not able to capture any detail of $\Psi(t, x)$ for $t<0$, unless the latter indeed vanish identically in practice.
It is apparent that the above equation falls back to that of the sourceless case by taking $S(s, x)=0$~\cite{agr-qnm-12}. 

The solution of the inhomogeneous differential equation Eq.~\eqref{master_eq_s} can be formally obtained by employing the Green function method.
To be specific,
\begin{eqnarray}
\hat{f}(s, x) = \int_{-\infty}^{\infty}G(s, x, x')(\mathcal{I}(s, x') - \mathscr{S}(s, x))dx' ,
\label{formal_solution_eq_s}
\end{eqnarray}
where the Green function satisfies
\begin{eqnarray}
G''(s, x, x')+\left(-s^2-V(x)\right)G(s, x, x')=\delta(x-x') .
\label{master_eq_Green}
\end{eqnarray}
It is straightforward to show that 
\begin{eqnarray}
G(s, x, x') = \frac{1}{W(s)}f_-(s, x_<)f_+(s, x_>) 
\label{master_eq_Green}
\end{eqnarray}
where $x_<\equiv \min(x, x')$, $x_>\equiv \max(x, x')$, and $W(s)$ is the Wronskian of $f_-$ and $f_+$.
Here $f_-$ and $f_+$ are the two linearly independent solutions of the corresponding homogeneous equation satisfying the physically appropriate boundary conditions~\cite{agr-qnm-12}
\begin{eqnarray}
\left\{
\begin{matrix}
f_-(s, x)\sim e^{s x} & \mathrm{as}\ x\to -\infty  \\ 
f_+(s, x)\sim e^{-s x} & \mathrm{as}\ x\to \infty
\end{matrix}\right.
\label{master_bc}
\end{eqnarray}
in asymptotically flat spacetimes, which are bounded with $\Re s>0$

The wave function thus can be obtained by evaluating the integral
\begin{eqnarray}
\Psi(t, x)=\frac{1}{2\pi i}\int_{\epsilon-i\infty}^{\epsilon+i\infty} e^{st}\hat{f}(s, x)ds ,
\label{inverse_Laplace}
\end{eqnarray}
where the integral is carried out on a vertical line in the complex plane, where $s=\epsilon+is_1$ with $\epsilon>0$.

Reminiscent of the toy model presented in the previous section, the discrete quasinormal frequencies are again established by evaluating Eq.~\eqref{inverse_Laplace} using the residue theorem.
In this case, one employs extended Jordan's lemma to close the contour with a large semicircle to the left of the original integration path~\cite{book-methods-mathematical-physics-03}.
The integration gives rise to the well-known result
\begin{eqnarray}
\oint e^{st}\hat{f}(s, x)ds = {2\pi i}\sum_q \mathrm{Res}\left(e^{st}\hat{f}(s, x), s_q\right) + (\mathrm{other\ contributions}),
\label{int_poles}
\end{eqnarray}
where $s_q$ indicates the poles inside the counter, 
``other contributions" are referring to those~\cite{agr-qnm-07, agr-qnm-08,agr-qnm-13,agr-qnm-14} from branch cut on the real axis, essential pole at the origin, and large semicircle.

Therefore, putting all pieces together, namely, Eqs.~\eqref{inverse_Laplace}, \eqref{int_poles}, \eqref{formal_solution_eq_s}, and \eqref{master_eq_Green} lead to
\begin{eqnarray}
\Psi(t, x)&=&\frac{1}{2\pi i}\int_{\epsilon-i\infty}^{\epsilon+i\infty} e^{st}G(s, x, x')\left[\mathcal{I}(s, x')-\mathscr{S}(s, x')\right]dx'ds \nonumber\\
&=&\frac{1}{2\pi i}\oint e^{st}\frac{1}{W(s)}\int_{-\infty}^{\infty}f_-(s, x_<)f_+(s, x_>)\left[\mathcal{I}(s, x')-\mathscr{S}(s, x')\right]dx'ds \nonumber\\
&=&\sum_q e^{s_qt}\mathrm{Res}\left(\frac{1}{W(s)}, s_q\right)\int_{-\infty}^{\infty}f_-(s_q, x_<)f_+(s_q, x_>)\left[\mathcal{I}(s_q, x')-\mathscr{S}(s_q, x')\right]dx' ,
\label{Laplace_eq_formal_solution}
\end{eqnarray}
where the residues are substituted after the last equality.
The above results can be rewritten as
\begin{eqnarray}
\Psi(t, x)=\sum_q c_q u_q(t, x) ,
\label{Laplace_eq_formal_solution_w_coefficients}
\end{eqnarray}
with
\begin{eqnarray}
c_q&=&\mathrm{Res}\left(\frac{1}{W(s)}, s_q\right)\int_{x_1}^{x_\mathcal{I}}f_-(s_q, x')\left[\mathcal{I}(s_q, x')-\mathscr{S}(s_q, x')\right]dx' \nonumber,\\
u_q(t, x)&=& e^{s_q t}f_+(s_q, x) ,
\label{Laplace_eq_formal_coefficients}
\end{eqnarray}
where one considers the case where the initial perturbations has a compact support, in other words, it locates in a finite range $x_1 < x' < x_\mathcal{I}$ and the observer is further to the right of it $x > x_\mathcal{I}$.

The quasinormal frequencies can be extracted from the temporal dependence of the solution, namely, Eq.~\eqref{Laplace_eq_formal_coefficients}.
Since $e^{i s_q t}$ is the only time-dependent factor, it is dictated by the values of the residues $s_q$.
The locations of the poles $s_q$ are entirely governed by the Green function Eq.~\eqref{master_eq_Green}, which, in turn, is determined by the zeros of the Wronskian.
Therefore, according to the formal solution Eq.~\eqref{Laplace_eq_formal_solution} or \eqref{Laplace_eq_formal_coefficients}, they are irrelevant to $c_q$, where the source $\mathscr{S}(s, x)$ is plugged in.
As $\Re s_q < 0$ the wave functions diverge at the spatial boundaries, which can be readily seen by substituting $s=s_q$ into Eqs.~\eqref{master_bc}, consistent with the results from the Fourier analysis~\cite{agr-qnm-review-02,agr-qnm-review-03,agr-qnm-lq-matrix-04}, as mentioned above.
It is observed that from Eq.~\eqref{Laplace_eq_formal_coefficients}, for given initial condition $\mathcal{I}(s_q, x)$, one may manipulate the external driving force $\mathscr{S}(s_q, x)$ so that only one single mode $s_q$ presented in the solution.

We note that the above discussions follow closely to those in the literature (see, for instance, Ref.~\cite{agr-qnm-12,agr-qnm-review-02}).
The only difference is that one subtracts the contribution of the external source, namely, $\mathscr{S}(s, x)$, from the initial condition $\mathcal{I}(s, x') $ in Eq.~\eqref{formal_solution_eq_s}.
It is well-known that the initial conditions of perturbation are irrelevant to the quasinormal frequencies, which characterize the {\it sound} of the black hole.
In this context, it is inviting to conclude that the external source term on the r.h.s. of the master equation Eq.~\eqref{master_eq} bears a similar physical content.
The Laplace formalism employed in this section facilitates the discussion.
On the other hand, from Eq.~\eqref{Laplace_eq_formal_coefficients}, one finds that the quasinormal modes' amplitudes will still be affected by the external source.
Overall, regarding the detection of quasinormal oscillations, the inclusion of external source does imply a significant modification of observables, such as the signal-noise ratio (SNR).
We note that the above discussions are valid under ths assumption that $\mathscr{S}(s, x)$ features a moderate spectrum in s-domain.
A notable exception will be discussed below in subsection C.

Before closing this subsection, we briefly comment on the equivalence between the above formalism based on Laplace transform and those in terms of Fourier analysis.
The results concerning the contour of integral and the quasinormal modes can be compared readily by taking $s=-i \omega$~\cite{agr-qnm-14}.  
To be more explicit, if one employs the Fourier transform together with the Green function method to solve Eq.~\eqref{master_eq_ns}, the formal solution has the form~\cite{book-methods-mathematical-physics-04}
\begin{eqnarray}
\Psi(t,x)=\int dx' G(t,x,x')\left.\frac{\partial \Psi(t,x')}{\partial t}\right|_{t=0}+\int dx' \frac{\partial G(t,x,x')}{\partial t}\left.\Psi(t,x')\right|_{t=0} ,
\label{solution_Green_Fourier}
\end{eqnarray}
where one considers the case without a source, and the contributions from the boundary at spatial infinity are irrelevant physically and have been ignored.
The Green function is the defined by 
\begin{eqnarray}
\frac{\partial^2}{\partial t^2}G(t,x,x')+\left(-\frac{\partial^2}{\partial x^2}+V\right)G(t,x,x')=\delta(t-t')\delta(x-x') ,
\label{Green_Fourier}
\end{eqnarray}

If we assume that the perturbations vanish identically for $t<0$, in other words, $G(t,x,x')=0$ for $t<0$.
By employing the Fourier transform in the place of Laplace transform, we have
\begin{eqnarray}
\tilde{G}(\omega, x,x')= \int_{-\infty}^\infty dt G(t,x,x') e^{i\omega t}=\int_0^\infty dt G(t,x,x') e^{i\omega t} .
\label{solution_Green_Fourier}
\end{eqnarray}
where $\tilde{G}(\omega, x,x')$ satisfies
\begin{eqnarray}
-\omega^2 \tilde{G}(\omega, x,x')+\left(-\frac{\partial^2}{\partial x^2}+V\right)\tilde{G}(\omega, x,x')=\delta(x-x') .
\label{master_equation_Green_Fourier}
\end{eqnarray}

Now, it is apparent that, up to an overall sign, the solution of Eq.~\eqref{master_equation_Green_Fourier} is essentially identical to Eq.~\eqref{master_eq_Green} in terms of Laplace transform.
The boundary contributions to the formal solution Eq.~\eqref{solution_Green_Fourier} are precisely those that the initial condition $\mathcal{I}$ contribute to Eq.~\eqref{formal_solution_eq_s}.
As discussed above, the main reason to employ the Laplace transform is that the formalism provides a transparent interpretation of the role taken by the external source.

\subsection{B. Kerr black hole metric}

As most black holes are likely to be rotating, calculations regarding stationary but rotating metrics are of potential significance from an experimental viewpoint.
In this subsection, we extend the above arguments to the case of the Kerr metric.
Here, the essential point is that the master equation of the Kerr metric cannot be rewritten in the form of a single second-order ordinary differential equation, such as Eq.~\eqref{master_eq}.
To be specific, by employing the method of separation of variables $\chi=e^{-i\omega t}e^{im\varphi}R(r)S(\theta)$, in standard Boyer-Lindquist coordinates, the master equation is found to be~\cite{agr-qnm-15}
 \bqn
 \lb{master_eq_Kerr}
\Delta^{-{\bar{s}}}\frac{d}{dr}\left(\Delta^{{\bar{s}}+1}\frac{d}{dr}\right)\hat{R}(\omega,r)+V\hat{R}(\omega,r)&=&0  ,\\
\left[\frac{d}{du}(1-u^2)\frac{d}{du}\right]{_{\bar{s}}S}_{\ell m}&& \nb\\
+\left[a^2 \omega^2u^2-2a\omega\bar{s}u+\bar{s}+{_{\bar{s}}A}_{\ell m}-\frac{(m+{\bar{s}}u)^2}{1-u^2}\right]{_{\bar{s}}S}_{\ell m}&=&0  \lb{master_eq_Kerr2},
 \eqn
where
 \bqn
 \lb{potential_Kerr}
V(r)&=&\frac{1}{\Delta(r)}\left\{(r^2+a^2)^2\omega^2-4Mam\omega r+a^2m^2+2ia(r-M)m{\bar{s}}-2iM(r^2-a^2){\bar{s}}\omega\right\}\nb\\
&+&(-a^2 \omega^2+2i\omega{\bar{s}}r-{_{\bar{s}}A}_{\ell m}) ,\nb\\
\Delta(r)&=&r^2-2Mr+a^2 ,\nb\\
u&\equiv&\cos\theta .
 \eqn
Also, $M$ and $aM\equiv J$ are the mass and angular momentum of the black hole, $m$ and $\bar{s}$ are the mass and spin of the perturbation field.
The solution of the angular part, ${_{\bar{s}}S}_{\ell m} = {_{\bar{s}}S}_{\ell m}(a\omega,\theta,\phi)$, is known as the spin-weighted spheroidal harmonics.
Here we have adopted the formalism in Fourier transform for simplicity.

Although both equations are ordinary differential equations, the radial equation for the quasinormal frequency $\omega$ now depends explicitly on ${_{\bar{s}}A}_{\ell m}$.
The latter is determined by the angular part of the master equation, which again involves $\omega$.
Therefore, when an external source is introduced, it seems one can no longer straightforwardly employ the arguments presented in the last section.
In particular, the arguments based on contour integral seem to work merely for the case where the radial equation is defined in such a way that it is independent of $\omega$. 
In what follows, however, we elaborate to show that the existing spectrum of quasinormal frequencies remains unchanged. 
We divide the proof into two parts.

The starting point is to assume that the solution of the homogeneous Eqs.~\eqref{master_eq_Kerr}-\eqref{master_eq_Kerr2} is already established.
First, let us focus on one particular quasinormal frequency $\omega = \omega_{n,\ell,m}$. 
For a given $\omega_{n,\ell,m}$, the angular part Eq.~\eqref{master_eq_Kerr2} is well-defined, and its solution is the spin-weighted spheroidal harmonics, ${_{\bar{s}}S}_{\ell m}$.
The latter is uniquely associated with a given value of ${_{\bar{s}}A}_{\ell m}$.
Now let us introduce an external source to the perturbation equations~\eqref{master_eq_Kerr}-\eqref{master_eq_Kerr2}.
One can show that $\omega=\omega_{n,\ell,m}$ must also be a pole of the Green function of the radial part of the resultant master equation.
The proof proceeds as follows.
It is known that the spin-weighted spheroid harmonics form a complete, orthogonal set for a given combination of $\bar{s}, a\omega$, and $m$~\cite{book-blackhole-physics-Frolov}.
Therefore, it can be employed to expand any arbitrary external source.
The expansion coefficient $\mathscr{S}$ is a function of radial coordinate $r$ will enter the radial part of the master equation, namely,
\bqn
\lb{master_eq_Kerr_source}
\Delta^{-{\bar{s}}}\frac{d}{dr}\left(\Delta^{{\bar{s}}+1}\frac{d}{dr}\right)\hat{R}(\omega,r)+V({_{\bar{s}}A}_{\ell m \omega_n})\hat{R}(\omega,r)=\mathscr{S}(\omega,r) ,
\eqn
while the angular part Eq.~\eqref{master_eq_Kerr2} remains the same.
It is note that ${_{\bar{s}}A}_{\ell m}$ is given with respect to given $\omega_{n,\ell,m}$, thus is denoted by ${_{\bar{s}}A}_{\ell m \omega_n}$.
Now, one is allowed to release and vary $\omega$ in Eq.~\eqref{master_eq_Kerr_source} in order to solve an equation similar to Eq.~\eqref{master_eq_s}.
As discussed in the last section, one may utilize the Green function method, namely, for real values of $\omega$ one solves
\bqn
\lb{master_eq_Kerr_Green}
\Delta^{-{\bar{s}}}\frac{d}{dr}\left(\Delta^{{\bar{s}}+1}\frac{d}{dr}\right)G(\omega,r,r')+V({_{\bar{s}}A}_{\ell m \omega_n})G(\omega, r,r')=\delta(r-r') ,
\eqn
and then considers analytic continuation of $\omega$ onto the complex plane.
It is evident that $\omega_{n,\ell,m}$ must be a pole of the above Green function.
This is because Eq.~\eqref{master_eq_Kerr_Green} does not involve the external source $S(\omega,r)$, and therefore the poles must be identical to the quasinormal frequencies of related sourceless scenario.
As we have already assumed, the latter, Eqs.~\eqref{master_eq_Kerr}-\eqref{master_eq_Kerr2}, have already be solved and $\omega_{n,\ell,m}$ is one of the quasinormal frequencies.
Besides, we note that the other poles of Eq.~\eqref{master_eq_Kerr_Green} are irrelevant, since they obviously do not satisfy Eq.~\eqref{master_eq_Kerr2}.
Moreover, the forms of Eq.~\eqref{master_eq_Kerr_source} as well as the Green function both change once a different value for $\omega_{n,\ell,m}$ is considered.

Secondly, let us consider a given $\omega$ but of arbitrary value.
Again, the angular part of the master equation Eq.~\eqref{master_eq_Kerr2} is a well-defined as an eigenvalue problem.
Subsequently, its solution, the spin-weighted spheroid harmonics, as a complete, orthogonal set for given $\bar{s}, a\omega$, and $m$, can be utilized to expand the external source.
One finds the following radial equation
\bqn
\lb{master_eq_Kerr_source_omega}
\Delta^{-{\bar{s}}}\frac{d}{dr}\left(\Delta^{{\bar{s}}+1}\frac{d}{dr}\right)\hat{R}(\omega,r)+V({_{\bar{s}}A}_{\ell m \omega})\hat{R}(\omega,r)=\mathscr{S}(\omega,r) .
\eqn
It is noted that the only difference is that ${_{\bar{s}}A}_{\ell m}$ explicitly depends on $\omega$ and it is therefore denoted as ${_{\bar{s}}A}_{\ell m \omega}$.
Although ${_{\bar{s}}A}_{\ell m \omega}$ is a function of $\omega$, the above equation is still a second order ordinary differential equation in $r$.
In other words, ${_{\bar{s}}A}_{\ell m \omega}$ can be simply viewed as a constant as long as one is solving the differential equation regarding $r$.
Once more, we will employ the Green function method, where the Green function in question satisfies
\bqn
\lb{master_eq_Kerr_Green_omega}
\Delta^{-{\bar{s}}}\frac{d}{dr}\left(\Delta^{{\bar{s}}+1}\frac{d}{dr}\right)G(\omega,r,r')+V({_{\bar{s}}A}_{\ell m \omega})G(\omega, r,r')=\delta(r-r') .
\eqn
Now, one is left to observe that the pole at $\omega=\omega_{n,\ell,m}$ of the Green function Eq.~\eqref{master_eq_Kerr_Green} is also a pole for the Green function Eq.~\eqref{master_eq_Kerr_Green_omega}.
The reason is that the pole $\omega=\omega_{n,\ell,m}$ of the Green function Eq.~\eqref{master_eq_Kerr_Green} corresponds to one of the zeros of the related Wronskian.
The latter is an algebraic (nonlinear) equation for $\omega$.
Likewise, the poles of the Green function Eq.~\eqref{master_eq_Kerr_Green_omega} also correspond to the zeros of a second Wronskian.
The latter is also an algebraic equation except that the constant ${_{\bar{s}}A}_{\ell m \omega_n}$ is replaced by ${_{\bar{s}}A}_{\ell m \omega}$, a function of $\omega$.
However, since 
\bqn
\left.{_{\bar{s}}A}_{\ell m \omega}\right|_{\omega=\omega_{n,\ell,m}}={_{\bar{s}}A}_{\ell m \omega_n}, \nb
\eqn
$\omega_{n,\ell,m}$ must also be a zero of the second Wronskian.
We, therefore, complete our proof that quasinormal frequencies $\omega=\omega_{n,\ell,m}$ are also the poles for the general problem with the external source.

The above results will be verified in the following section against explicit numerical calculations. 
Moreover, we note that additional poles, besides those originated from the zeros of the Wronskian, might also be introduced owing to the presence of an external source.
One interesting example is that they may come from the ``quasi-singularity" of the external source.
This possibility will be explored in the next section.

\section{IV. Additional modes introduced by the external source}

In the above, we mentioned that when a sinusoidal external force is applied, the frequency of the steady-state oscillation is known to be identical to that of the driving force.
Moreover, it is understood that the resonance takes place when the magnitude of the driving frequency matches that of the natural frequency of the oscillator.
At a first glimpse, since the driven force's frequency is usually independent of the natural frequency of the oscillator, the above results seem to contradict our conclusion so far.
In previous sections, we have shown that if the external source is not singular, namely, characterized by a moderate frequency spectrum, the system's natural frequencies will not be affected.
However, the results given in Eqs.~\eqref{toy_sol_source_res} and \eqref{Laplace_eq_formal_solution} will suffer potential modification when the source term $\mathscr{S}$ contains singularity.

First of all, we argue that in the context of black hole physics, the sinusoidal driving force is not physically relevant, as it corresponds to some perpetual external energy source.
A physically meaningful scenario should be related to some dissipative process, such as when the external source is characterized by some resonance state.
In particular, the resonance will be associated with a complex frequency, where the imaginary part of the frequency gives rise to the half-life of the resonance decay.
Mathematically, the external source thus possesses a pole on the complex plane.
The physical requirement of dissipative nature indicates that, in the Laplace s-domain, the real part of $s=-i\omega$ is negative.
In other words, the poles of the source term, if any, must be located on the left of the imaginary axis, and therefore they are inside the contour in Eq.~\eqref{int_poles}.
In turn, according to the residue theorem, they will introduce additional quasinormal frequencies to the temporal oscillations.

In the case of the toy model, if a given frequency governs the driven force, it corresponds to the case where a single frequency dominates $ \mathscr{S}(\omega, n)$, namely, $\mathscr{S}(\omega, n)\sim \delta(\omega - \omega_R)$\footnote{It is noted that the Dirac delta function has to be viewed as a limit of a sequence of complex analytic functions, such as the Poisson kernel, for the discussions carried out in terms of the contour integral to be valid.}.
Regarding Eq.~\eqref{toy_sol_source_res}, this will affect the evaluation of residue.
To be specific, the driving force gives rise to a pole in the complex plane at $\omega = \omega_R - i\epsilon$, where the additional infinitesimal imaginary part $i\epsilon$ corresponds to a resonance state with infinite half-life.
As a result, the long-term steady-state oscillations will be entirely overwhelmed by the contribution from this pole.
In other words, a {\it normal} mode will govern the system's late-time behavior, consistent with our initial observations.
As discussed above, in the case of the black hole, one deals with some external resonance source, which corresponds to quasinormal modes.
Since the external driving force is independent of the nature of the system, those quasinormal modes are not determined by the Green function Eq.~\eqref{master_eq_Green}.
In other words, by definition, they are not governed by the black hole parameters as for the conventional quasinormal modes of the metric.
In the following section, we will show numerically that additional quasinormal frequency can indeed be introduced by the external source.

It is worth noting that the physical nature of external source discussed in the present section is different from that of the initial pulse or initial condition.
To be specific, in literature~\cite{agr-qnm-12, agr-qnm-review-02, agr-qnm-review-03}, the quainormal modes are defined regarding the perturbation equation in the time domain, Eq.~\eqref{master_eq_ns}.
By considering the Laplace transform, the equation is rewritten where the initial condition $\mathcal{I}(s, x)$ appears on the r.h.s. as a source term.
As discussed above, this term might affect the amplitudes of the quasinormal oscillation but is irrelevant to the quasinormal frequencies.
This is because the real physical content it carries is an initial pulse.
It is evident that a harmonic oscillator's initial condition will never affect the oscillator's natural frequency.
On the other hand, as in Eq.~\eqref{master_eq}, if one introduces a source term directly onto the r.s.h. of the master equation in the time domain, one might encounter a different scenario.
As discussed above, now the physical content resides in the well-known example that a driven harmonic oscillator will follow the external force's frequency when, for instance, a sinusoidal driving force is applied.
Therefore, these are two {\it distinct} scenarios associated with the term external source, which, as discussed in the text, lead to different implications.
The above conclusion can be confirmed mathematically.
In fact, it can be readily shown that the term $\mathcal{I}(s, x)$ given by the Laplace transform must not contain any singularity.
Observing Eq.~\eqref{master_eq_sic}, its frequency dependence is linear in $s$, thus averting any potential pole on the complex plane.

\section{V. Numerical results}

In this section, we demonstrate that the results obtained analytically in previous sections agree with numerical calculations.
To be specific, we first solve the inhomogeneous differential equations numerically.
Subsequently, the evolution of perturbations in the time domain is used to extract the dominant complex frequencies by utilizing the Prony method.
These frequencies are then compared against the numerical results of the corresponding quasinormal modes, obtained by standard approaches.

We first demonstrate the precision of our numerical scheme by studying the toy model presented in section II.
Then we proceed to show the results for the Schwarzschild as well as Kerr metrics.

For the toy model, one considers the master Eq.~\eqref{toy_eq} numerically for $\tau=1, L=1, n=1$ and the source Eq.~\eqref{toy_source} where
\begin{eqnarray}
\mathscr{S}_{(1)}(\omega, n)=1 ,
\label{toy_source_num_01}
\end{eqnarray}
and
\begin{eqnarray}
\mathscr{S}_{(2)}(\omega, n)=\frac{1}{\omega^2+1} ,
\label{toy_source_num_02}
\end{eqnarray}

\begin{figure}
\begin{tabular}{cc}
\vspace{0pt}
\begin{minipage}{225pt}
\centerline{\includegraphics[width=200pt]{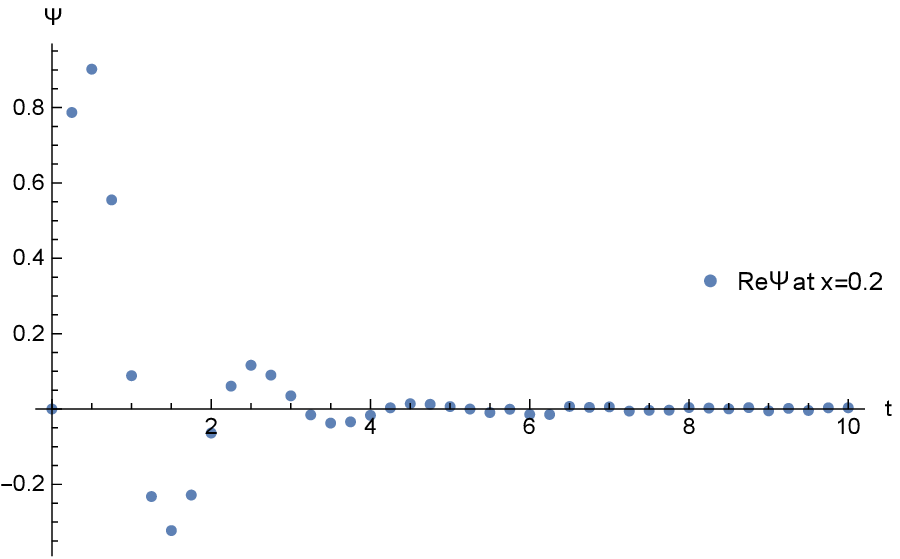}}
\end{minipage}
&
\begin{minipage}{225pt}
\centerline{\includegraphics[width=200pt]{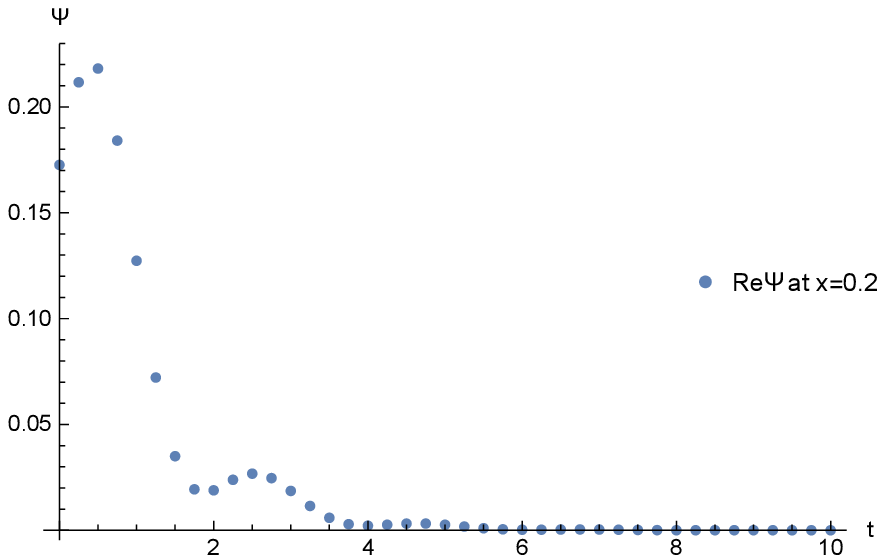}}
\end{minipage}
\end{tabular}
\renewcommand{\figurename}{Fig.}
\caption{(Color online)
The calculated time series of the toy model for two different types of sources given in Eq.~\eqref{toy_source_num_01} and~\eqref{toy_source_num_02} are shown in the left and right plot, respectively.
The calculations are carried out to generate a total of 40 points in the time series.}
\label{evolution_toy}
\end{figure}

Our first goal is to find the temporal dependence of the solution for the two arbitrary sources chosen above.
This is accomplished by first solve Eq.~\eqref{toy_eq} in the frequency space and then carry out an inverse Fourier transform at an arbitrary given position $x$ for various time instant $t$.
Although part of the above procedure can be obtained analytically, we have chosen to adopt the numerical approach, since later on, for more complicated scenarios, we will eventually resort to the ``brutal" numerical force.
The resultant time series are shown in Fig.~\ref{evolution_toy}.
It is observed that the temporal evolution indeed follows the pattern of quasinormal oscillations.

In order to extract the quasinormal frequencies, the Prony method~\cite{agr-qnm-16} is employed.
The method is a powerful tool in data analysis and signal processing.
It can be used to extract the complex frequencies from a regularly spaced time series.
The method is implemented by turning a non-linear minimization problem into that of linear least squares in matrix form.
As shown below, in practice, even a small dataset of 40 points is often sufficient to extract precise results.
In the following, we choose the modified least-squares Prony~\cite{agr-qnm-16} over others, as the impact of noise is not significant in our study.

For Eq.~\eqref{toy_source_num_01}, the two most dominating quasinormal frequencies are found to be $\omega_{(1)}^{\pm}=-0.999i-2.982, -0.999i+2.967$.
For Eq.~\eqref{toy_source_num_02}, one also obtains two dominating complex frequencies $\omega_{(2)}^{\pm}=-0.999999i-2.978190, -0.999998i+2.978189$.
The numerical results together with their respective weights are shown in Tab.~\ref{PronyList}.
When compared with the analytic values $\omega^{\pm}=-i\pm \sqrt{\pi^2-1}\sim -i\pm 2.978188$, one finds that desired precision has been achieved.

Next, one proceeds to the case of the Schwarzschild black hole.
Here, we consider massless scalar perturbation with the following source term 
\begin{eqnarray}
\mathscr{S}_{(3)}(\omega, x)=\frac{1}{1+\omega^2}  \frac{1}{rf^2(r)} V(r)e^{i\omega r} ,
\label{Schwarzschild_source_num_03}
\end{eqnarray}
where we take $\bar{s}=0, r_h=2M=1, \ell=1, L=\infty$, $V$ and $f$ are given by Eq.~\eqref{V_master}-\eqref{f_master}, the tortoise coordinate $x=\int dr/f$.
It is noted that the factor $e^{i\omega r}V(r)/f^2(r)$ is introduced to guarantee that the source satisfies appropriate boundary conditions.
The remaining factor $\frac{1}{1+\omega^2}\frac{1}{r}$ can largely be chosen arbitrarily.

\begin{figure}
\begin{tabular}{cc}
\vspace{0pt}
\begin{minipage}{225pt}
\centerline{\includegraphics[width=200pt]{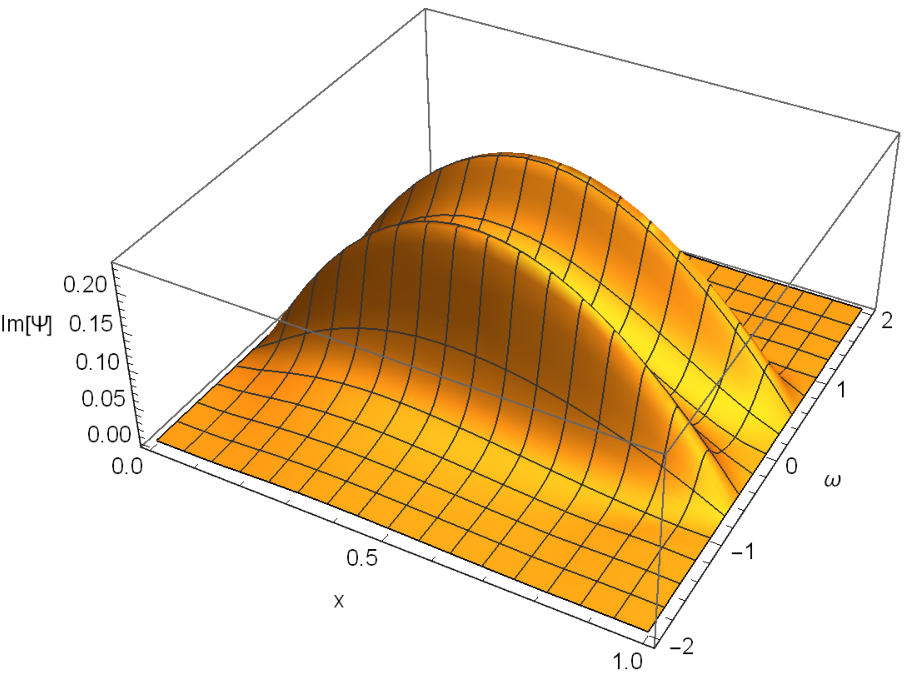}}
\end{minipage}
&
\begin{minipage}{225pt}
\centerline{\includegraphics[width=200pt]{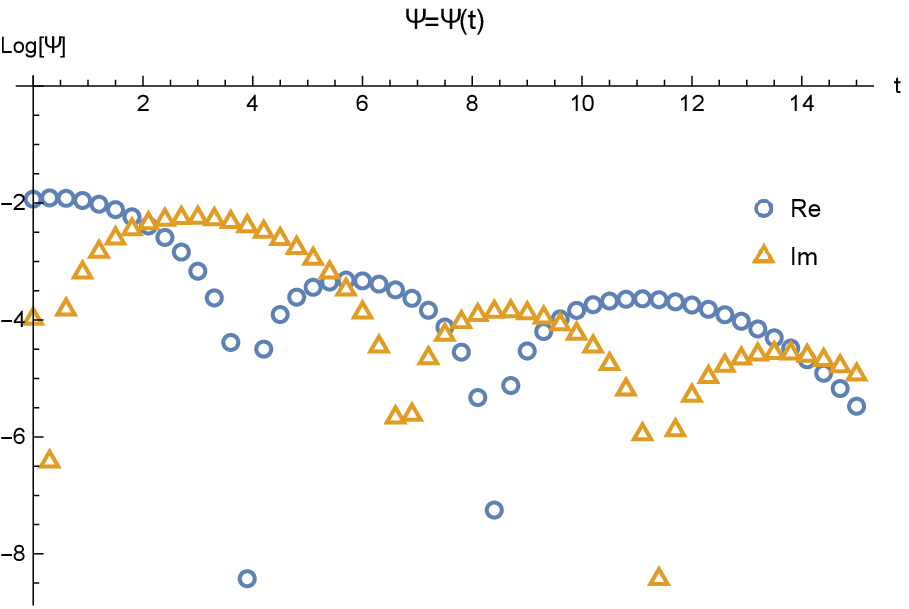}}
\end{minipage}
\end{tabular}
\renewcommand{\figurename}{Fig.}
\caption{(Color online)
Results on massless scalar perturbations in Schwarzschild black hole metric with external source.
Left: The calculated imaginary part of the numerical solution of the master equation in the frequency domain, shown as a 2D function of $\omega$ and $x$.
Right: The calculated time series of the massless scalar perturbations.
The calculations are carried out to generate a total of 50 points in the time series.}
\label{evolution_Schwarzschild}
\end{figure}

To find the temporal evolution, we again solve the master equation in the frequency domain of Eq.~\eqref{master_eq} by employing a adapted matrix method~\cite{agr-qnm-lq-matrix-01,agr-qnm-lq-matrix-02}.
To be specific, the radius coordinate is transform into a finite interval $x\in [0,1]$ by $r\to \frac{2M}{1-x}$, which subsequently discretized into 22 spatial grids.
For simplicity, we consider $\alpha=1, \ell=1$.
By expressing the function and its derivatives in terms of the function values on the grids, the differential equation is transformed into a system of linear equations represented by a matrix equation.
The solution of the equation is then obtained by reverting the matrix, as shown in the left plot of Fig.~\ref{evolution_Schwarzschild}.
Subsequently, the inverse Fourier transform is carried out numerically at a given spatial grid $x=\frac{5}{21}$, presented in the right plot of Fig.~\ref{evolution_Schwarzschild}.
As an approximation, the numerical integration is only carried for the range $\omega\in [-20,20]$, where a necessary precision check has been performed.
By employing the Prony method, one can readily extract the most dominate quasinormal frequency.
The resultant value is $\omega_{(3)}=-0.5847 - 0.1954 i$, consistent with $\omega_{n=0,\ell=1}=-0.5858 - 0.1953 i$ obtained by the matrix method~\cite{agr-qnm-lq-matrix-02}.

Now, we are ready to explore the master equation Eq.~\eqref{master_eq_Kerr_source} for Kerr metric with the following form for the source term
\begin{eqnarray}
\mathscr{S}_{(4)}(\omega, r)=\frac{1}{1+\omega^2}\frac{r(r-r_+)}{\Delta} e^{i\omega r}  ,
\label{Kerr_source_num_04}
\end{eqnarray}
where $r_+=M+\sqrt{M^2-a^2}$ is the radius of the event horizon.
Here, the form $\frac{r(r-r_+)}{\Delta} e^{i\omega r}$ is to guarantee that the external source vanishes at the spatial boundary as $a\to 0$, so that the asymptotical behavior of the wave function remains unchanged.
Also, the factor $\frac{1}{1+\omega^2}$ is again introduced, based on the observation that its presence in Eq.~\eqref{toy_source_num_02} has led to better numerical precision.
The latter is probably due to that the resultant numerical integration regarding the inverse Fourier transform converges faster.
This choice turns out to be particularly helpful in the present scenario where the numerical precision becomes an impeding factor.
In the following calculations, we choose $M=0.5, a=0.3$, and $\ell=2$.

\begin{figure}
\begin{tabular}{cc}
\vspace{0pt}
\begin{minipage}{225pt}
\centerline{\includegraphics[width=200pt]{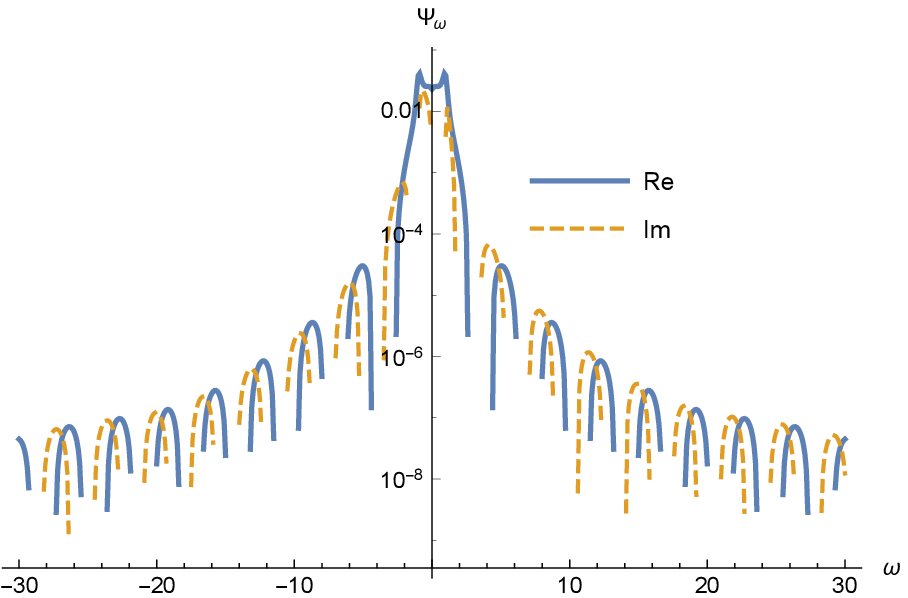}}
\end{minipage}
&
\begin{minipage}{225pt}
\centerline{\includegraphics[width=200pt]{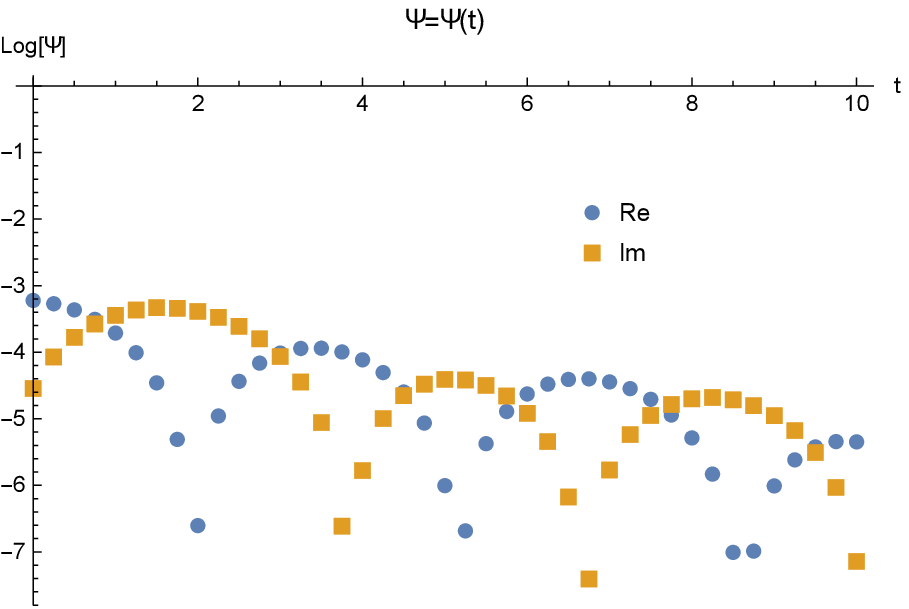}}
\end{minipage}
\end{tabular}
\renewcommand{\figurename}{Fig.}
\caption{(Color online)
Results on massless scalar perturbations in Kerr black hole metric with external source.
Left: The real and imaginary parts of the master equation's numerical solution in the frequency domain, evaluated at $x=\frac{4}{21}$.
Right: The calculated time series of the massless scalar perturbations.
The calculations are carried out to generate a total of 40 points in the time series.}
\label{evolution_Kerr}
\end{figure}

Based on the matrix method, the entire range of the spatial and polar coordinates $r$ and $\theta$ is divided by 22 grids.
Subsequently, the radial, as well as angular parts of the master equation, are discretized into two matrix equations~\cite{agr-qnm-lq-matrix-03}.
We first solve the angular part of the master equation Eq.~\eqref{master_eq_Kerr2} for a given $\omega$ to obtain ${_{\bar{s}}A}_{\ell m \omega}$.
This can be achieved with relatively high precision, namely, with a {\it WorkingPrecision} of $100$ in {\it Mathematica}.
The obtained $\omega$ to obtain ${_{\bar{s}}A}_{\ell m \omega}$ is substituted back into Eq.~\eqref{master_eq_Kerr_source} to solve for the wave function in the frequency domain.
To improve efficiency, we only carry out the calculation for a given spatial point at $x=\frac{4}{21}$, without losing generality.
The resultant wave function is shown in the left plot of Fig.~\ref{evolution_Kerr}.
To proceed, we evaluate the wave function at $600$ discrete points between $-30 <\omega < 30$ and then use those values to approximate the numerical integration in the frequency domain.
The resultant time series with 40 points are shown in the right plot of Fig.~\ref{evolution_Kerr}.
By using the Prony method, the most dominant quasinormal frequency is found to be $\omega_{(4)}=-0.9981 - 0.1831 i$, in good agreement with the value $\omega_{n=0,\ell=2}=0.9918 - 0.1869 i$ obtained by the 21th order matrix method~\cite{agr-qnm-lq-matrix-03}.

\begin{table}[htb]
\begin{center}
\scalebox{1.00}{\begin{tabular}{|c|c|c|c|c|}
\hline
\multirow{2}{*}{source} & \multicolumn{2}{c|}{$1$st} & \multicolumn{2}{c|}{$2$nd} \\ \cline{2-5}  
& \multicolumn{1}{c|}{~$\omega^-$~} & \multicolumn{1}{c|}{weight} & \multicolumn{1}{c|}{~$\omega^+$~} & \multicolumn{1}{c|}{weight} \\ \hline
$\mathscr{S}_{(1)}$& $ -0.999878i -2.982507$ & $7.5\times 10^{-01}$ & $-0.998883i  +2.966696$ & $7.5\times 10^{-01}$    \\  \hline
$\mathscr{S}_{(2)}$ & $-0.999999i-2.978190$ & $7.0\times 10^{-02}$  & $-0.999998i+2.978189$ & $7.0\times 10^{-02}$   \\  \hline
\multirow{2}{*}{} & \multicolumn{2}{c|}{$3$rd} & \multicolumn{2}{c|}{$4$th} \\ \cline{2-5}  
& \multicolumn{1}{c|}{~$\omega$~} & \multicolumn{1}{c|}{weight} & \multicolumn{1}{c|}{~$\omega$~} & \multicolumn{1}{c|}{weight} \\ \hline
$\mathscr{S}_{(1)} $& $-2.236222i -7.901108$ & $4.7\times 10^{-03}$ & $-2.229535 i-11.873706$ & $2.2\times 10^{-03}$   \\  \hline
$\mathscr{S}_{(2)}$ & $-0.999999i -6.035914\times 10^{-08}$ & $2.5\times 10^{-01} $ & $0.016891i -9.887078$ & $2.1\times 10^{-07}$  \\  \hline
\end{tabular}}
\end{center}
\caption{The calculated quasinormal frequencies by using the Prony method for the source terms Eqs.~\eqref{toy_source_num_01} and \eqref{toy_source_num_02}.
The numerical code has been implemented to extract five modes while the first four most dominante ones, as well as their respective amplitudes, are listed.}\label{PronyList}
\end{table}

Last but not least, we investigate whether the poles in the external source will also demonstrate itself in the resultant temporal series.
This can be demonstrated by revisiting the toy model.
In particular, it is evident the external source Eq.~\eqref{toy_source_num_02} contains two poles on the complex plane, for $t>0$ the relevant pole is $\omega^e=-i$.
Therefore, if everything checks out, the additional frequency $\omega^e$ must also be captured by the Prony method.
Taking a close look at the results listed in Tab.~\ref{PronyList} reveals that this is indeed the case.
For the source term $\mathscr{S}_{(1)} $, the first two modes overwhelm others by two orders of magnitude.
On the other hand, concerning $\mathscr{S}_{(2)} $, not only it helps to improve the precision of the numerical integration, a third dominant mode appears, which reads $\omega^e_{(2)}=-0.999999i -6.035914\times 10^{-08}$.
It readily confirmed that the poles in the driving force are relevant, and present themselves as additional quasinormal modes in the resultant time series.

One can proceed to show explicitly that it is also the case in the context of black hole configurations.
However, on the numerical aspect, it is a bit tricky.
We note that, by comparing Eq.~\eqref{toy_source_num_02} against Eq.~\eqref{Schwarzschild_source_num_03}, it is evident that the latter also contains the pole at $\omega^e$.
Unfortunately, the present numerical scheme is not robust enough to pick out this singularity.
In order to accomplish our goal, one might deliberately bring the singularities to the region where their detection becomes feasible while the frequency domain integral still converges reasonably fast.
This can be achieved by replacing the source term in Eq.~\eqref{Schwarzschild_source_num_03} by an appropriately chosen form
\begin{eqnarray}
\mathscr{S}_{(3)}(\omega, x)=\frac{1}{(\omega+\frac13 i+1)(\omega-\frac13 i+1)}  \frac{1}{rf^2(r)} V(r)e^{i\omega r} .
\label{Schwarzschild_source_num_04}
\end{eqnarray}
It gives rise to an additional pair of singularities, out of which $\omega^{e-}=-\frac13 i- 1$ is relevant to the contour in question.
By carrying out an identical procedure, we manage to extract the latter using the present algorithm..
The first two dominant modes extracted by the Prony method are found to be $\omega_{(5)}=-0.5824 - 0.1896 i$ and $\omega_{(6)}=-0.9952 - 0.3326 i$.
In other words, both the fundamental quasinormal mode and the singularity in the source term are identified successfully.
We are looking forward to improving the algorithm further so that its application to more sophisticated scenarios becomes viable.

\section{VI. Further discussions and concluding remarks}

To summarize, in this work, we study the properties of external sources in blackhole perturbations.
We show that even with the presence of the source term in the time-domain, the quasinormal frequencies may largely remain unchanged.
In this case, the physical content of the external source is an initial pulse.
The statement is valid for various types of perturbation in both static and/or stationary metrics.
Although, for rotating black holes, the arguments are elaborated with additional subtlety.
We also discuss the physically relevant scenraio where the external source acts as a driving force and introduces additional modes.
The findings are then attested against the numerical calculations for several particular scenarios.

It is noted that in our discussions, the effects of the branch cut on the negative real axis have not been considered.
These discontinuity from the branch cut arises from that of the solution of the homogeneous radial equation, which satisfies the boundary condition at infinity.
As a result, their effects remain unchanged as the external source is introduced.
Moreover, as the branch cut stretches from the origin, it primarily associated with the late-time behavior of the perturbations.
Therefore, they are largely not relevant to the quasinormal frequencies in the context of the present study.

The numerical calculations carried out in the present paper only involve rather straightforward scenarios such as the Schwarzschild metric.
Since our results are expected to be valid in a more general context, as mentioned above, it is physically meaningful to explore further the possible implications in more sophisticated cases.
These include the perturbations in modified gravity theories, such as the scalar-tensor theories.
One relevant feature of the theory is that the scalar perturbations are entirely decoupled from those of the Einstein tensor.
In some recent studies, the metric perturbations in the DHOST theory are found to possess a source term~\cite{agr-modified-gravity-dhost-07, agr-modified-gravity-dhost-08}.
Besides, the master equation for scalar perturbations is shown to be a first-order differential equation decoupled from the Einstein tensor perturbations.
Subsequently, for such specific cases, one may obtain the general solution (see, for example, Eq.~(26) of Ref.~\cite{agr-modified-gravity-dhost-08}), which does not contain any pole in the frequency domain.
In other words, the discussions in section III.B can be readily applied to these cases. 
In this regard, we have demonstrated that while the magnitude of the perturbation wave function is tailored by the source and initial condition, the quasinormal frequencies might stay the same. 
Therefore, the findings of the present work seem to indicate a subtlety in extracting information on the stealth scalar hair in the DHOST theory via quasinormal modes.
In our view, it is rather inviting to explore the details further, and also for other modified theories of gravity.
Further studies along this direction are in progress.

\section*{Acknowledgments}
WLQ is thankful for the hospitality of Chongqing University of Posts and Telecommunications. 
We gratefully acknowledge the financial support from
Funda\c{c}\~ao de Amparo \`a Pesquisa do Estado de S\~ao Paulo (FAPESP),
Funda\c{c}\~ao de Amparo \`a Pesquisa do Estado do Rio de Janeiro (FAPERJ),
Conselho Nacional de Desenvolvimento Cient\'{\i}fico e Tecnol\'ogico (CNPq),
Coordena\c{c}\~ao de Aperfei\c{c}oamento de Pessoal de N\'ivel Superior (CAPES),
and National Natural Science Foundation of China (NNSFC) under contract Nos. 11805166, 11775036, and 11675139.
A part of the work was developed under the project INCTFNA Proc. No. 464898/2014-5.
This research is also supported by the Center for Scientific Computing (NCC/GridUNESP) of the S\~ao Paulo State University (UNESP).

\bibliographystyle{h-physrev}
\bibliography{references_qian}

\end{document}